\documentclass[fleqn,usenatbib]{mnras}

\usepackage{newtxtext,newtxmath}

\usepackage[T1]{fontenc}

\DeclareRobustCommand{\VAN}[3]{#2}
\let\VANthebibliography\thebibliography
\def\thebibliography{\DeclareRobustCommand{\VAN}[3]{##3}\VANthebibliography}

\usepackage{graphicx}	

\usepackage{amsmath}	
\usepackage{amssymb}	
\usepackage[normalem]{ulem}
\usepackage{booktabs} 

\usepackage{algorithmic,algorithm}

\newcommand{\orcid}[1]{\href{https://orcid.org/#1}{\includegraphics[width=10pt]{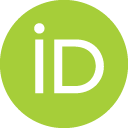}}}

\newcommand{\Algo}[1]{Algorithm~\ref{algo:#1}}

\newcommand{\algolabel}[1]{\label{algo:#1}}

\title[Accelerating inference with PMC]{Accelerating astronomical and cosmological inference with Preconditioned Monte Carlo}

\author[M. Karamanis et al.]{
Minas Karamanis \orcid{0000-0001-9489-4612},$^{1}$\thanks{E-mail: minas.karamanis@ed.ac.uk}
Florian Beutler \orcid{0000-0003-0467-5438},$^{1}$
John A. Peacock \orcid{0000-0002-1168-8299},$^{1}$
David Nabergoj \orcid{0000-0001-6882-627X}$^{2}$
and Uro\v{s} Seljak \orcid{0000-0003-2262-356X}$^{3}$
\\
$^{1}$Institute for Astronomy, University of Edinburgh, Royal Observatory, Blackford Hill, Edinburgh EH9 3HJ, UK\\
$^{2}$Faculty of Computer and Information Science, University of Ljubljana, Ve\v{c}na pot 113, 1000 Ljubljana, Slovenia \\
$^{3}$Physics Department, University of California and Lawrence Berkeley National Laboratory Berkeley, CA 94720, USA
}

\date{Accepted XXX. Received YYY; in original form ZZZ}

\pubyear{2022}

\begin{document}
\label{firstpage}
\pagerange{\pageref{firstpage}--\pageref{lastpage}}
\maketitle

\begin{abstract}
    We introduce \emph{Preconditioned Monte Carlo} (PMC), a novel Monte Carlo method for Bayesian inference that facilitates efficient sampling of probability distributions with non--trivial geometry. PMC utilises a \emph{Normalising Flow} (NF) in order to decorrelate the parameters of the distribution and then proceeds by sampling from the preconditioned target distribution using an adaptive \emph{Sequential Monte Carlo} (SMC) scheme. The results produced by PMC include samples from the posterior distribution and an estimate of the model evidence that can be used for parameter inference and model comparison respectively. The aforementioned framework has been thoroughly tested in a variety of challenging target distributions achieving state--of--the--art sampling performance. In the cases of \emph{primordial feature} analysis and \emph{gravitational wave} inference, PMC is approximately $50$ and $25$ times faster respectively than \emph{Nested Sampling} (NS). We found that in higher dimensional applications the acceleration is even greater. Finally, PMC is directly parallelisable, manifesting linear scaling up to thousands of CPUs. An open--source \texttt{Python} implementation of PMC, called \texttt{pocoMC}, is publicly available at \url{https://github.com/minaskar/pocomc}.
\end{abstract}




\section{Introduction}
\label{sec:intro}

Modern astronomical and cosmological analyses have largely adopted the framework of \emph{Bayesian probability} for tasks of parameter inference and model comparison. In the Bayesian context, the \emph{posterior probability distribution} $\mathcal{P}(\theta)=P(\theta |\mathcal{D}, \mathcal{M})$, meaning the probability distribution of the parameters $\theta$ of a model $\mathcal{M}$, given some data $\mathcal{D}$ and the model $\mathcal{M}$ is given by Bayes' theorem:
\begin{equation}
    \mathcal{P}(\theta) = \frac{\mathcal{L}(\theta)\pi(\theta)}{\mathcal{Z}}\,,
\end{equation}
where $\mathcal{L}(\theta) = P(\mathcal{D}|\theta, \mathcal{M})$ is the \emph{likelihood function}, $\pi(\theta)=P(\theta |\mathcal{M})$ is the \emph{prior} probability distribution, and $\mathcal{Z} = P(\mathcal{D}|\mathcal{M})$ is the \emph{model evidence} or \emph{marginal likelihood} that acts as a normalisation constant for the posterior probability distribution. For a detailed introduction to Bayesian probability theory we refer the reader to~\citet{jaynes2003probability, gregory2005bayesian, mackay2003information} and the reviews~\citet{trotta2017bayesian, sharma2017markov} for its use in astronomy and cosmology.

In tasks of parameter inference, the goal is to infer the values of physical and nuisance parameters from the data along with the respective uncertainties. Mathematically, this is formulated as the problem of estimating expectation values (e.g. mean values, standard deviations, 1--D and 2--D marginal posterior distributions, etc.) that correspond to high--dimensional integrals over the posterior probability density. During the past two decades, \emph{Markov chain Monte Carlo} (MCMC) has been established as the standard computational tool for the calculation of such integrals (see e.g. ~\citep{speagle2019conceptual} for a review). 
MCMC methods generate a sequence of correlated samples, called a Markov chain, that are distributed according to the posterior probability distribution. Those samples can then be used in order to numerically estimate expectation values. Examples of MCMC software implementations in the astronomical and cosmological community are \textit{emcee}~\citep{foreman2013emcee} and \textit{zeus}~\citep{karamanis2021zeus}.

Most modern MCMC methods are based upon the \emph{Metropolis--Hastings} (MH) paradigm that consists of two steps~\citep{metropolis1953equation, hastings1970}. In the first step, known as the \emph{proposal step}, a new sample is drawn from a known proposal distribution that depends only on the position of the current sample. The validity of the new sample, and thus the decision on whether to add it or not to the Markov chain, is determined in the second step, known as the \emph{acceptance step}, which takes into account the new sample, the old sample (i.e. current state) and the proposal distribution that was used in order to generate it. Arguably, the most important element of an efficient MCMC method is the choice of proposal distribution. The degree to which the proposal distribution characterises the local geometry of the target distribution determines the sampling efficiency (i.e. the rate of effectively independent samples) of the method. Unfortunately, choosing  or tuning the optimal proposal distribution for a given target distribution is not an easy task. However, certain optimal proposal distributions are known for specific classes of target distributions. For instance, in the case of a normal or Gaussian target distribution, using a normal proposal distribution of the form $\mathcal{N}(\theta,2.38^{2}\Sigma / D)$, where $\Sigma$ is the covariance matrix of the target density, $\theta$ is the current state of the chain, and $D$ is the number of dimensions yields the maximum sampling efficiency scheme with acceptance rate of $23.4\%$ in the acceptance step of MH~\citep{gelman1997weak}. Alternatively, one can use a simpler proposal distribution of the form $\mathcal{N}(u,1)$ where $u=f(\theta)$ and $f$ is a suitable transformation. In this case, $f(\theta)$ is proportional to $L^{-1}\theta$ where $L$ is the lower triangular matrix of the \emph{Cholesky decomposition} of the covariance matrix $\Sigma = L L^{T}$. In other words, assuming that a suitable transformation can be found, one can increase the sampling efficiency of an MCMC method. This notion of preconditioning is central for the discussion that will follow in the next section.

In recent years, the need for higher sampling efficiency when the correlations between parameters are strong enough or the posterior exhibits multiple modes, as well as the required computation of the model evidence $\mathcal{Z}$ for model comparison tasks, motivated the development of more advanced sampling methodologies and algorithms. One very popular approach is the \emph{Sequential Monte Carlo} (SMC) algorithm~\citep{del2006sequential}, which evolves a set of particles through a series of intermediate steps that bridge the gap between the prior distribution and the posterior distribution by geometrically interpolating between them. Another class of algorithms called \emph{Nested Sampling} (NS)~\citep{skilling2004nested} attempts to approach the problem of Bayesian computation from a slightly different perspective. Instead of evolving a set of particles though a series of geometrically--interpolated steps between prior and posterior distribution, NS splits the posterior distribution into many slices and attempts to sample each slice individually with an appropriate weighting scheme. Many popular versions and implementations of NS exist in the astronomical literature~\citep{speagle2020dynesty, buchner2021ultranest, handley2015polychord, feroz2009multinest}. Whereas both SMC and NS largely addressed the problem of multimodality, the performance of both methods is still very sensitive to the geometry of the target distribution, meaning the presence of strong non--linear correlations.

In this paper, we introduce \emph{Preconditioned Monte Carlo} (PMC), a novel Monte Carlo method for Bayesian inference that extends the range of applications of SMC to target distributions with non--trivial geometry, strong non--linear correlations between parameters, and severe multimodality. PMC achieves this by first preconditioning, or transforming the geometry of the target distribution into a more manageable one using a generative model known as a \emph{Normalising Flow} (NF)~\citep{papamakarios2021normalizing}, before sampling using a SMC scheme. \citet{hoffman2019neutra} used a NF to neutralise the bad geometry in Hamiltonian Monte Carlo (HMC)~\citep{betancourt2017conceptual} achieving great results in terms of sampling speed but unreliable estimates for unknown target distributions. \citet{moss2020accelerated} used a NF in order to parameterise efficient MCMC proposals and used it in the context of NS achieving a substantial speedup on several challenging distributions. Both of the aforementioned works used NFs as preconditioning transformations, the first in the context of HMC and the second in NS. In the context of NS and SMC, NFs have also been used as a sampling component of the algorithm~\citep{albergo2019flow,williams2021nested, arbel2021annealed}, albeit not as a preconditioner but as a density from which new samples can be generated independently. The novelty of our work lies in the use of NFs as preconditioning transformations in the context of SMC, thus achieving both robustness and high sampling efficiency.

The structure of the rest of the paper is the following: Section \ref{sec:method} consists of a detailed presentation of the method, Section \ref{sec:tests} includes a wide range of empirical tests that act as a demonstration of PMC's sampling performance, and Section \ref{sec:conclusions} is reserved for the conclusions.

We also release a \texttt{Python} implementation of PMC, called \texttt{pocoMC}, which is publically available at \url{https://github.com/minaskar/pocomc} and detailed documentation with installation instructions and examples at \url{https://pocomc.readthedocs.io}. The code implementation is described in the accompanying paper~\citep{karamanis2022pocomc}.

\section{Method}
\label{sec:method}

\subsection{Sequential Monte Carlo}

In this subsection, we will present a brief introduction to SMC algorithms. For a more detailed exposition, we refer the reader to~\citet{naesseth2019elements}. We begin by first introducing the concept of \emph{importance sampling}, which is crucial for understanding the function of SMC. Assuming that we have a target probability density $\pi(\theta)$ that we are able to evaluate up to an unknown multiplicative constant, then if we define another density $\rho(\theta)$, called the \emph{importance sampling density}, such that $\rho(\theta) =0 \Rightarrow \pi(\theta) = 0$ then the following relation holds for any expectation value:
\begin{equation}
    \label{eq:importance_sampling}
    \begin{split}
        \mathrm{E}_{p}[f(\theta)] & = \int f(\theta) w(\theta) \rho(\theta) d\theta \Big/ \int w(\theta) \rho(\theta) d\theta \\
        & = \mathrm{E}_{\rho}[f(\theta)w(\theta)]/\mathrm{E}_{\rho}[w(\theta)]\, ,
    \end{split}
\end{equation}
for any function $f(\theta)$ where $w(\theta) = p(\theta) / \rho(\theta)$ are called importance weights. We can use samples from the importance density $\rho(\theta)$ in order to estimate the above expectation value without explicitly sampling from the target density $p(\theta)$.

A common measure of the quality of using the importance sampling density $\rho(\theta)$ to approximate $p(\theta)$ is the \emph{Effective Sample Size}, defined as:
\begin{equation}
    \label{eq:effective_sample_size}
    \text{ESS} = \mathrm{E}_{\rho}[w(\theta)]^{2} / \mathrm{E}_{\rho}[w(\theta)^{2}]\,.
\end{equation}
Unfortunately, in high--dimensional scenarios it is difficult to find an appropriate importance sampling density that ensures that the ESS is high enough for the variance of the expectation value to be low. This is exactly the problem that SMC methods address.

SMC samplers extend the importance sampling procedure from the setting of two densities (i.e. importance sampling density and target density) to a sequence of $T$ probability distributions $\lbrace p_{t}\rbrace _{t=1}^{T}$ in which each individual density $p_{t}$ acts as the importance density for the next one in the series. The method proceeds by pushing a collection of $N$ particles $\lbrace \theta_{t}^{k}\rbrace _{k=1}^{N}$ through this sequence of densities until the last one is reached. Each iteration of an SMC algorithm consists of three main steps:
\begin{enumerate}
    \item \textbf{Mutation} -- The population of particles is moved from $\lbrace \theta_{t-1}^{k}\rbrace _{k=1}^{N}$ to $\lbrace \theta_{t}^{k}\rbrace _{k=1}^{N}$ using a \emph{Markov transition kernel} $K_{t}(\theta '|\theta)$ that defines the next importance sampling density
    \begin{equation}
        \label{eq:Markov_kernel}
        p_{t}(\theta ') = \int p_{t-1}(\theta) K_{t}(\theta '|\theta) d\theta \, .
    \end{equation}
    In practice, this step consists of running multiple short MCMC chains (i.e. one for each particle) to get the new states $\theta '$ starting from the old ones $\theta$.
    
    \item \textbf{Correction} -- The particles are reweighted according to the next density in the sequence. This step consists of multiplying the current weight $W_{t}^{k}$ of each particle by the appropriate importance weight:
    \begin{equation}
        \label{eq:importance_weight}
        w_{t}(\theta_{t}) = p_{t}(\theta_{t-1})/p_{t-1}(\theta_{t-1})\,.
    \end{equation}
    
    \item \textbf{Selection} -- The particles are resampled according to their weights which are then set to $1/N$. This can be done using \emph{multinomial resampling} or more advanced schemes. The purpose of this step is to eliminate particles with low weight and multiply the ones with high weights.
\end{enumerate}

An important feature of SMC is that it allows for the unbiased estimation of the ratios of normalising constants
\begin{equation}
    \label{eq:normalising_ratios}
    \mathcal{Z}_{t}/\mathcal{Z}_{t-1} = \sum_{k=1}^{N}W_{t-1}^{k}w_{t}(\theta_{t-1}^{k})\,,
\end{equation}
between subsequent densities. This is of paramount importance in cases in which the first density in the series corresponds to the prior distribution (i.e. with $\mathcal{Z}=1$) and the last to the posterior distribution. Then, SMC methods can be used in order to compute the model evidence $\mathcal{Z}$ for tasks of model comparison.

In principle, there are arbitrary many ways to construct the sequence of densities $\lbrace p_{t}\rbrace _{t=1}^{T}$.  A very common way to do so is to geometrically interpolate between two densities $\rho(\theta)$ and $p(\theta)$:
\begin{equation}
    \label{eq:temperature_annealing}
    p_{t}(\theta) \propto \rho(\theta)^{1-\beta_{t}}p(\theta)^{\beta_{t}},\quad t = 1, \dots, T\,
\end{equation}
parameterised by a \emph{temperature annealing} ladder:
\begin{equation}
    \label{eq:temperature_ladder}
    \beta_{1} = 0 < \beta_{2} < \dots < \beta_{T} = 1\,.
\end{equation}
In the Bayesian context, a natural choice of geometric interpolation is from the prior $\pi(\theta)$ to the posterior:
\begin{equation}
    \label{eq:temperature_annealing_bayesian}
    p_{t}(\theta) \propto \pi(\theta)\mathcal{L}(\theta)^{\beta_{t}},\quad t = 1, \dots, T\,
\end{equation}
where $\mathcal{L}(\theta)$ is the likelihood function. In practice, it can still be difficult to choose a good temperature schedule. However, this can be done adaptively by selecting the next value of $\beta_{t}$ such that the ESS is a constant $\alpha$ fraction of the number of particles $N$. Numerically, this can be done by solving
\begin{equation}
    \label{eq:bisection_method}
    \bigg( \sum_{k=1}^{N}w_{t+1}^{k}(\beta_{t+1})\bigg)^{2}\Big/ \sum_{k=1}^{N}w_{t+1}^{k}(\beta_{t+1})^{2}=\alpha N\,,
\end{equation}
the next $\beta_{t+1}$ such that $\beta_{t}<\beta_{t+1}\leq 1$ using, for instance, the \emph{bisection method}.

\begin{figure}
    \centering
	\centerline{\includegraphics[scale=0.70]{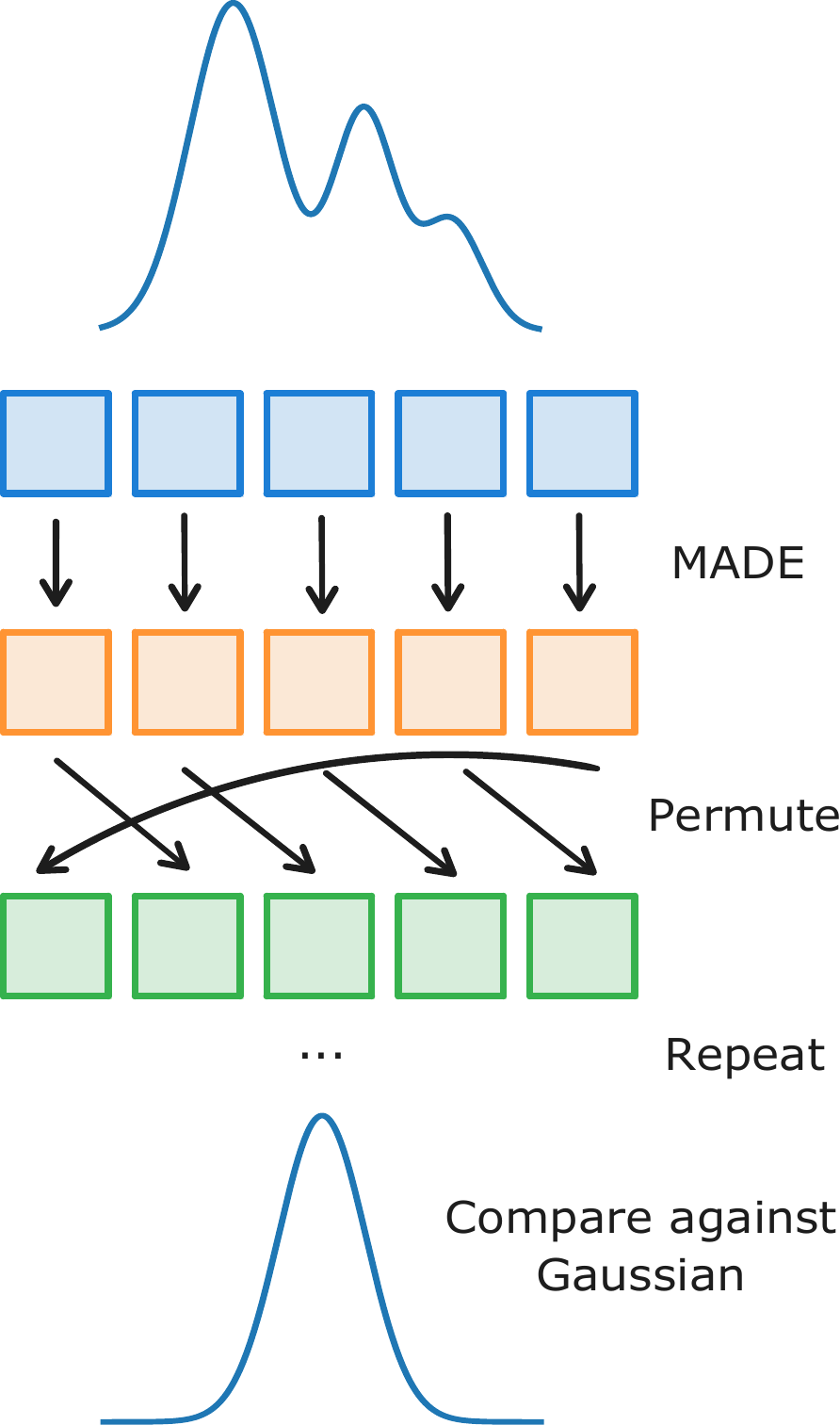}}
    \caption{Illustration of the inference scheme of a \emph{Masked Autoregressive Flow} (MAF). The arrows show the conditional dependence of the variables as well as the action of the \emph{Masked Autoregressive Density Estimation} (MADE) layer. The input target probability density (top) is mapped into a multivariate normal distribution (bottom). A sequence of MADE layers and permutations is repeated multiple times in order to increase the flexibility of the flow.}
    \label{fig:maf}
\end{figure}

\subsection{Normalising Flows}

Normalising flows (NF) are generative models, which can facilitate efficient and exact density estimation~\citep{papamakarios2021normalizing}. They are based on the formula of change--of--variables $\theta = f(u)$ where $u$ is sampled from a base distribution $u \sim p_{u}(u)$ (i.e. usually a normal distribution). The NF is a bijective mapping between the base distribution $p_{u}(u)$ and the often more complex target distribution $p_{\theta}(\theta)$ that can be evaluated exactly using
\begin{equation}
    \label{eq:jacobian_transform}
    p_{\theta}(\theta) = p_{u}(f^{-1}(\theta))\bigg|\det\bigg( \frac{\partial f^{-1}}{\partial \theta}\bigg) \bigg| \,,
\end{equation}
where the Jacobian determinant is tractable.

NFs are usually parameterised by neural networks. However, neural networks are not invertible in general, and the Jacobian is not generally tractable. Special care needs to be taken when choosing the architecture of the neural network to ensure the invertability of the transformation and the tractability of the Jacobian. For instance, if the forward transformation is $\theta_{i} = u_{i}\exp(\alpha_{i})+\mu_{i}$ and inverse transformation is $u_{i} = (\theta_{i} -  \mu_{i})\exp(-\alpha_{i})$, where $\mu_{i}$ and $\alpha_{i}$ are constants, then it is straightforward to show that the Jacobian satisfies
\begin{equation}
    \label{eq:jacobian_example}
    \bigg|\det\bigg( \frac{\partial f^{-1}}{\partial \theta}\bigg) \bigg| = \exp \bigg( -\sum_{i}\alpha_{i}\bigg)\,.
\end{equation}

To this end, we chose to use the \emph{Masked Autoregressive Flow} (MAF), which has been used many times successfully for density estimation tasks due to its superior performance and high flexibility compared to alternative models~\citep{papamakarios2017masked}. A MAF consists of many stacked layers of a simpler generative model, called \emph{Masked Autoregressive Density Estimator} (MADE)~\citep{germain2015made}, with subsequent permutations of its outputs as shown in Figure \ref{fig:maf}. A MADE model decomposes a joint density $p(\theta)$ as a product of conditionals $p(\theta) = \prod_{i}p(\theta_{i}|\theta_{1:i-1})$ that ensures that any given value $\theta_{i}$ is only a function of the previous values thus maintaining the \emph{autoregressive property}. When the MADE is based on an \emph{autoencoder}, then \emph{masking} is required in order to remove connections between different units in different layers, so as to preserve the aforementioned autoregressive property.

\subsection{Preconditioning}

\begin{figure*}
    \centering
	\centerline{\includegraphics[scale=0.58]{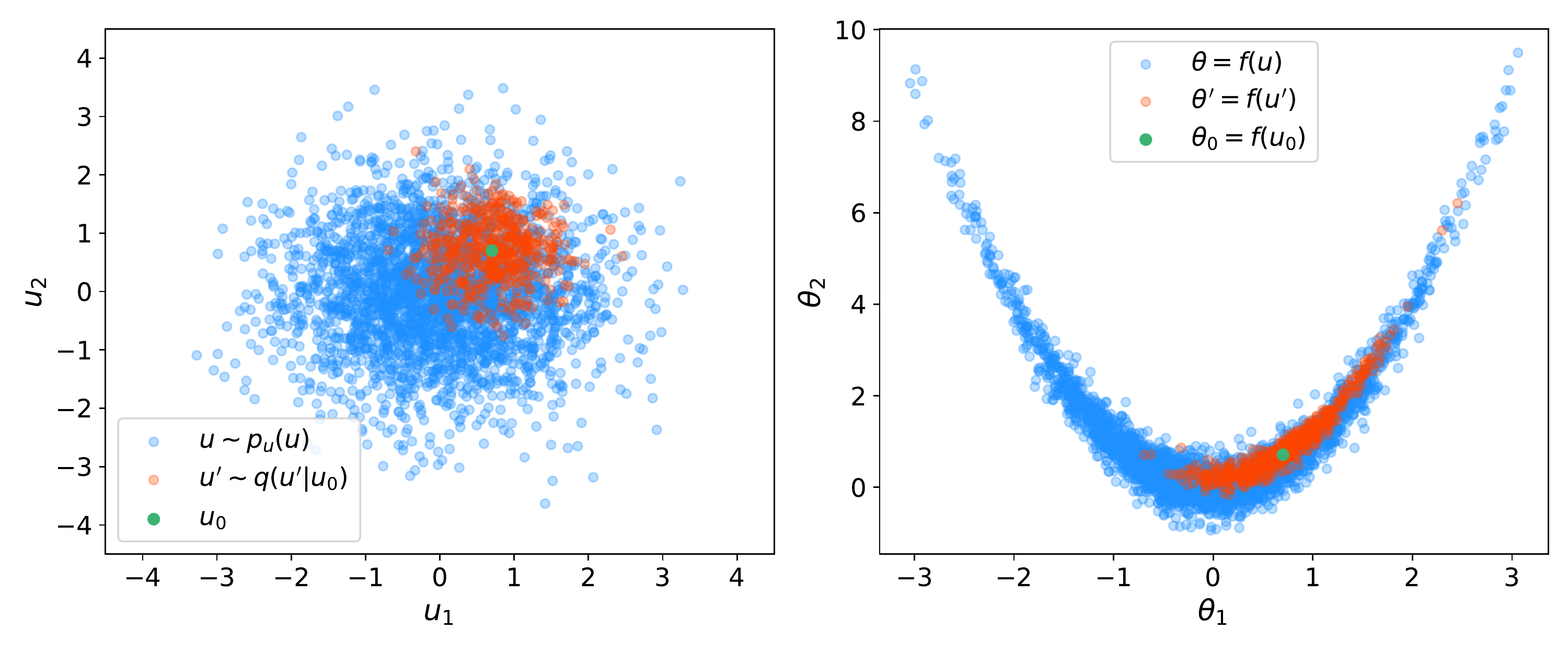}}
    \caption{The figure illustrates the effect of preconditioning on the \emph{Rosenbrock} distribution. The right panel shows samples (blue) from the true correlated distribution and the left panel shows samples (blue) from the preconditioned/transformed one. The orange samples in the left panel are drawn from a symmetric normal proposal distribution centred around the green point $u_{0}$ and they correspond to the respective orange points in the right panel. In other words, the transformed samples from the simple proposal in the left panel correspond to samples that capture the local geometry of the true target distribution in the right panel.}
    \label{fig:proposal}
\end{figure*}

Most MCMC methods struggle to sample efficiently from highly correlated or skewed target distributions. Often, transforming the parameters of the distribution before sampling, a process also known as \emph{preconditioning}, using appropriate change--of--variable transformations, can help ameliorate this effect by disentangling the dependence between parameters. This is equivalent to choosing an appropriate proposal distribution in the context of \emph{Metropolis--Hastings} (MH) methods. However, finding a valid transformation and selecting an appropriate proposal distribution is often difficult a priori; and there is no obvious way of making this joint choice in an optimal way. For instance, a linear transformation $\theta \leftarrow L^{-1}\theta$ where $L$ is the lower triangular matrix of the \emph{Cholesky decomposition} of the \emph{sample covariance} matrix $\Sigma = LL^{T}$ can remove only linear correlations and is not effective against non--linear ones. More sophisticated transformations, such as the use of the \emph{chirp mass} and \emph{mass ratio} instead of the individual black--hole masses in gravitational wave astronomy requires expert knowledge that is problem--specific.

The \emph{Metropolis acceptance criterion} employed by MH methods in order to maintain detailed balance is
\begin{equation}
    \label{eq:Metropolis_criterion}
    \alpha = \min \bigg( 1, \frac{p_{\theta}(\theta ')q(\theta | \theta ')}{p_{\theta }(\theta)q(\theta' | \theta)}\bigg)\,,
\end{equation}
where $p_{\theta}(\theta)$ is the target distribution and $q(\theta' | \theta)$ is the proposal distribution. For a general transformation $\theta = f(u)$ and its inverse $u = f^{-1}(\theta)$ the modified \emph{Metropolis acceptance criterion} takes the following form
\begin{equation}
    \label{eq:Metropolis_criterion_transform}
    \alpha = \min \left( 1, \frac{p_{\theta}(f^{-1}(u'))q(u | u ')\Big|\det \frac{\partial f^{-1}(u')}{\partial u'}\Big|}{p_{\theta }(f^{-1}(u))q(u ' | u)\Big|\det \frac{\partial f^{-1}(u)}{\partial u}\Big|}\right)\,,
\end{equation}
where the Jacobian determinant also appears. In this formulation of MH, the sampler samples the distribution in the transformed space and then samples are pushed through the $\theta = f(u)$ transformation to the original space. Assuming that the transformation $\theta = f(u)$ induces a simpler geometry onto the transformed space, sampling using the above acceptance criterion can be substantially more efficient. 

Figure \ref{fig:proposal} shows one such transformation that transforms the banana--shaped \emph{Rosenbrock} distribution into a unit--variance normal distribution and \emph{vice versa}. The same figure also demonstrates the effectiveness of simple proposal distributions $q(u'|u)$ in the transformed space. A symmetric normal proposal distribution $q(u'|u_{0})$ centred around a point $u_{0}$ corresponds to a highly effective proposal distribution in the original space, which captures the local geometry of the target distribution around that point.

\subsection{Preconditioned Monte Carlo}

\emph{Preconditioned Monte Carlo} (PMC) is the result of the amalgamation of SMC, NFs and preconditioning as they were introduced in the previous paragraphs. In particular, we suggest the use of the transformation $\theta = f(u)$ of a NF in order to precondition the \emph{Mutation} step of SMC. A pseudocode of the algorithm is presented at \Algo{pmc}. The \emph{Mutation} step in this case consists of $N$ \emph{Random--Walk Metropolis} (RWM) steps, meaning MH with an isotropic Gaussian proposal distribution centred around the current state of the Markov chain, in which the algorithm targets the preconditioned density. We fix the acceptance rate of MH to its optimal value $23.4\%$ between temperature steps by adapting the proposal scale~\citep{gelman1997weak}. As the optimal proposal scale of MH for a Gaussian target distribution is
\begin{equation}
    \label{eq:optimal_scale}
    \sigma_{\textrm{opt}} = \frac{2.38}{\sqrt{D}}\,,
\end{equation}
where $D$ is the number of dimensions, we can assess the performance of the NF preconditioner by estimating the ratio of the true scale $\sigma$ to the optimal one $\sigma_{\textrm{opt}}$. Assuming that the NF preconditions perfectly the target density and maps it into a unit--variance Gaussian distribution, this ratio should be equal to one. In practice, this ratio can deviate slightly from the optimal value of unity, and one can utilise this ratio as a metric of the preconditioning quality. The number $N$ of the MCMC steps performed in each iteration is determined adaptively during the run. The process we used is based on the mean correlation coefficient between the initial positions of the particles in the beginning of an iteration and their current positions. In particular, the particles are updated using MCMC until their mean correlation coefficient drops below a prespecified threshold. The lower the threshold, the higher the number $N$ of MCMC steps. It is important to note that the correlation coefficient is computed in the preconditioned $u$ space.

\begin{algorithm}
\caption{Preconditioned Monte Carlo}
    \algolabel{pmc}
\begin{algorithmic}[1]
\STATE{\textbf{input} Number of particles $N$}
\STATE{$t\leftarrow 1$, $\beta_{1}\leftarrow 0$, $\mathcal{Z}\leftarrow 1$}
\STATE{\textbf{for} $k=1$ \textbf{to} N \textbf{do} sample $\theta_{1}^{k}\sim \pi(\theta)$ and set $W_{1}^{k}=1/N$}
\STATE{train $\theta = f(u)$ using $\lbrace \theta_{1}^{k}\rbrace _{k=1}^{N}$}
\WHILE{$\beta_{t}\neq 1$}
    \STATE{$t\leftarrow t + 1$}
    \STATE{$\beta_{t}\leftarrow$ solution to Eq. \ref{eq:bisection_method}}
    \STATE{\textbf{for} $k=1$ \textbf{to} N \textbf{do} $w_{t}^{k}\leftarrow W_{t-1}^{k}\mathcal{L}(\theta)^{\beta_{t}-\beta_{t-1}}$}
    \STATE{$\mathcal{Z}\leftarrow \mathcal{Z}\sum_{k=1}^{N}w_{t}^{k}$}
    \STATE{$\lbrace \Tilde{\theta}_{t-1}^{k}\rbrace _{k=1}^{N} \leftarrow$ resample $\lbrace \theta_{t-1}^{k}\rbrace _{k=1}^{N}$ according to $\lbrace w_{t}^{k}\rbrace _{k=1}^{N}$}
    \STATE{\textbf{for} $k=1$ \textbf{to} N \textbf{do} $W_{t}^{k}\leftarrow 1/N$}
    \STATE{$\lbrace \theta_{t}^{k}\rbrace _{k=1}^{N}\leftarrow$ move $\lbrace \Tilde{\theta}_{t-1}^{k}\rbrace _{k=1}^{N}$ according to $K_{t}\left( \lbrace \theta_{t}^{k}\rbrace _{k=1}^{N}\leftarrow \lbrace \Tilde{\theta}_{t-1}^{k}\rbrace _{k=1}^{N}\, ; f \right)$}
    \STATE{train $\theta = f(u)$ using $\lbrace \theta_{t}^{k}\rbrace _{k=1}^{N}$}
\ENDWHILE
\STATE{\textbf{return} samples $\lbrace \theta_{t}^{k}\rbrace _{k=1}^{N}$ and estimate of the marginal likelihood $\mathcal{Z}$}
\end{algorithmic}
\end{algorithm}

\subsection{Hyperparameters}

We organise the hyperparameters of PMC into two groups, those related to the normalising flow and those related to the SMC algorithm. The first group consists of structure and training hyperparamaters for the NF. The NF structure parameters include the number of MADE layers (\texttt{blocks}), as well as the number of neurons per hidden layer (\texttt{neurons}). The NF training hyperparameters include the learning rate (\texttt{lr}) of the Adam optimiser~\citep{kingma2014adam}, the maximum number of epochs (\texttt{epochs}), the training batch size (\texttt{batch}), the tolerance for early stopping (\texttt{tolerance}), and the Laplace prior scale (\texttt{b}) used for regularisation. On the other hand, the SMC hyperparameters include the number of particles (\texttt{particles}), the desired effective sample size (\texttt{ESS}), and the correlation coefficient threshold (\texttt{threshold}). The default values for those hyperparameters are shown in Table \ref{tab:default}. We found that this configuration was robust and efficient for a wide range of applications and thus decided to recommend it as the default choice.

\begin{table}
    \centering
    \caption{The table shows the default values for the hyperparameters of PMC.}
    \def\arraystretch{1.1}
    \begin{tabular}{lc|lc}
        \toprule[0.75pt]
        \multicolumn{2}{l}{NF hyperparameters} & \multicolumn{2}{l}{SMC hyperparameters} \\
        \midrule[0.5pt]
        \texttt{blocks} & $6$ & \texttt{particles} & $1000 - 4000$ \\
        \texttt{neurons} & $3\times D$ & \texttt{ESS} & $95\%$ \\
        \texttt{batch} & $1000$ & \texttt{threshold} & $75\%$ \\
        \texttt{epochs} & $500$ &   &   \\
        \texttt{tolerance} & $30$ &   &   \\
        \texttt{lr} & $10^{-2} - 10^{-5}$ &  &  \\
        \texttt{b} & $0.2$ &  &  \\
        \bottomrule[0.75pt]
    \end{tabular}
    \label{tab:default}
\end{table}

\subsection{Parallelization}

\begin{figure}
    \centering
	\centerline{\includegraphics[scale=0.45]{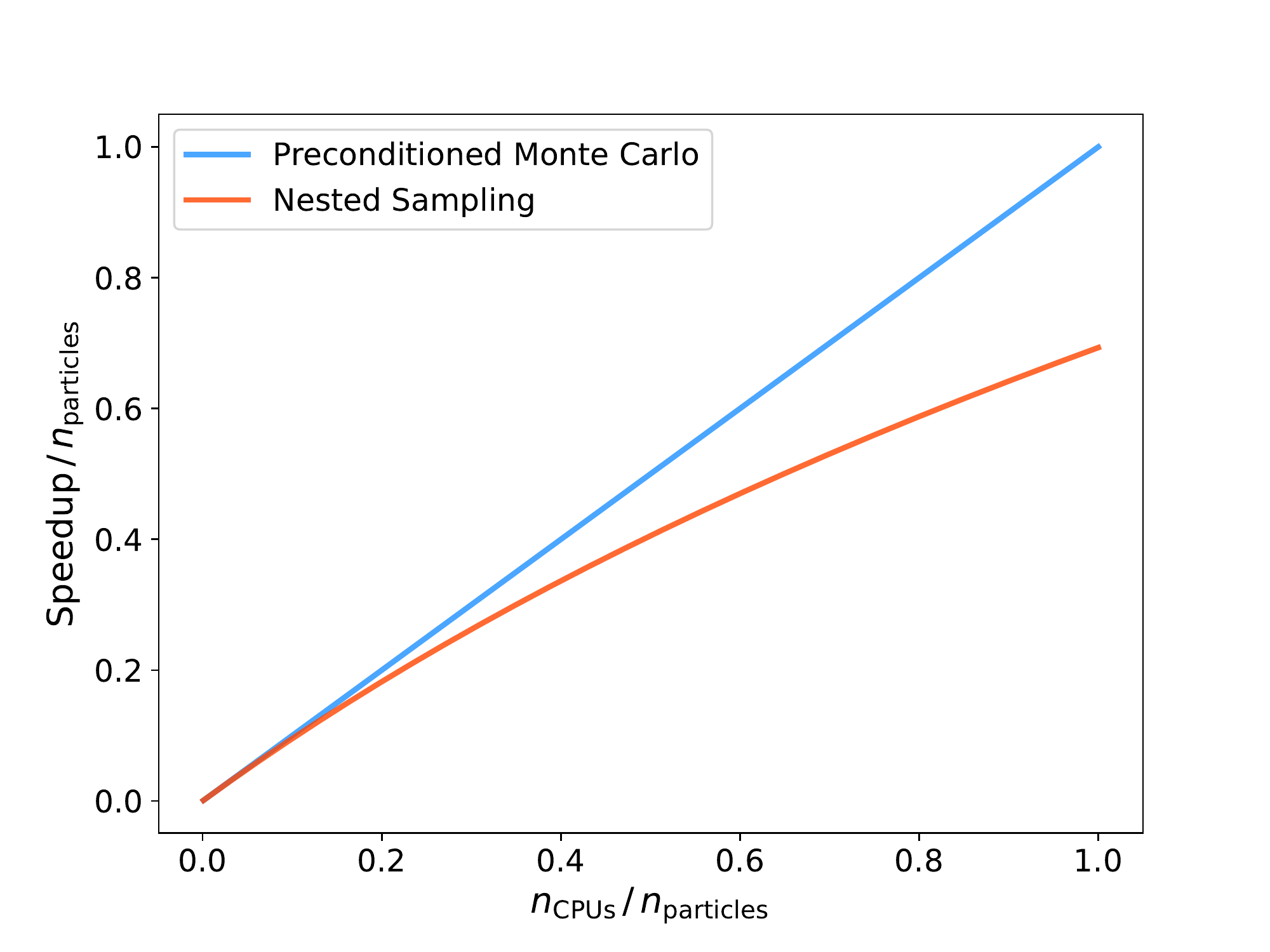}}
    \caption{Parallelization of PMC compared to nested sampling. PMC (blue) exhibits linear speedup compared to the sub--linear one achieved by NS (orange).}
    \label{fig:speedup}
\end{figure}

An important property of PMC is its ideal scaling with the available number of CPUs. In particular, the mutation step of PMC is exactly parallelisable, meaning that the speedup gained by using more than one CPU scales linearly with the number of CPUs as long as $n_{\mathrm{CPUs}} \leq n_{\mathrm{particles}}$. Similar methods that also use a large collection of particles scale less favourably. For instance, \emph{Nested Sampling} (NS) exhibits sub--linear scaling as shown in Figure \ref{fig:speedup} of \cite{handley2015polychord}. The aforementioned scaling characteristic of PMC renders it ideal for computationally costly applications that are often encountered in astronomy and cosmology.

\section{Empirical Evaluation}
\label{sec:tests}

In this section, we present two toy examples and two realistic parameter inference examples that reproduce common astronomical and cosmological analyses. In all cases, the hyperparameters of PMC were set to their default values as shown in Table \ref{tab:default}. In both analyses, the performance of PMC is compared to that of SMC using the same settings (e.g. number of particles, ESS, etc.) as PMC but no preconditioning, as well as \emph{Nested Sampling} (NS), a popular particle Monte Carlo alternative~\footnote{We used the popular \texttt{Python} implementation \texttt{dynesty}~\citep{speagle2020dynesty} for NS.}. The metric that we use in order to evaluate the performance of each method is the total number of model evaluations performed until convergence. Convergence in all methods is well--defined: in PMC and SMC the algorithm converges when $\beta=1$, whereas in NS the run stops when less than $1\%$ of the model evidence is left unaccounted.  All other computational costs are negligible, including the training and evaluation of the normalising flow in the case of PMC that only required a few seconds for the whole inference procedure. All methods used $1000$ particles.

\subsection{Rosenbrock distribution}

\begin{figure}
    \centering
	\centerline{\includegraphics[scale=0.42]{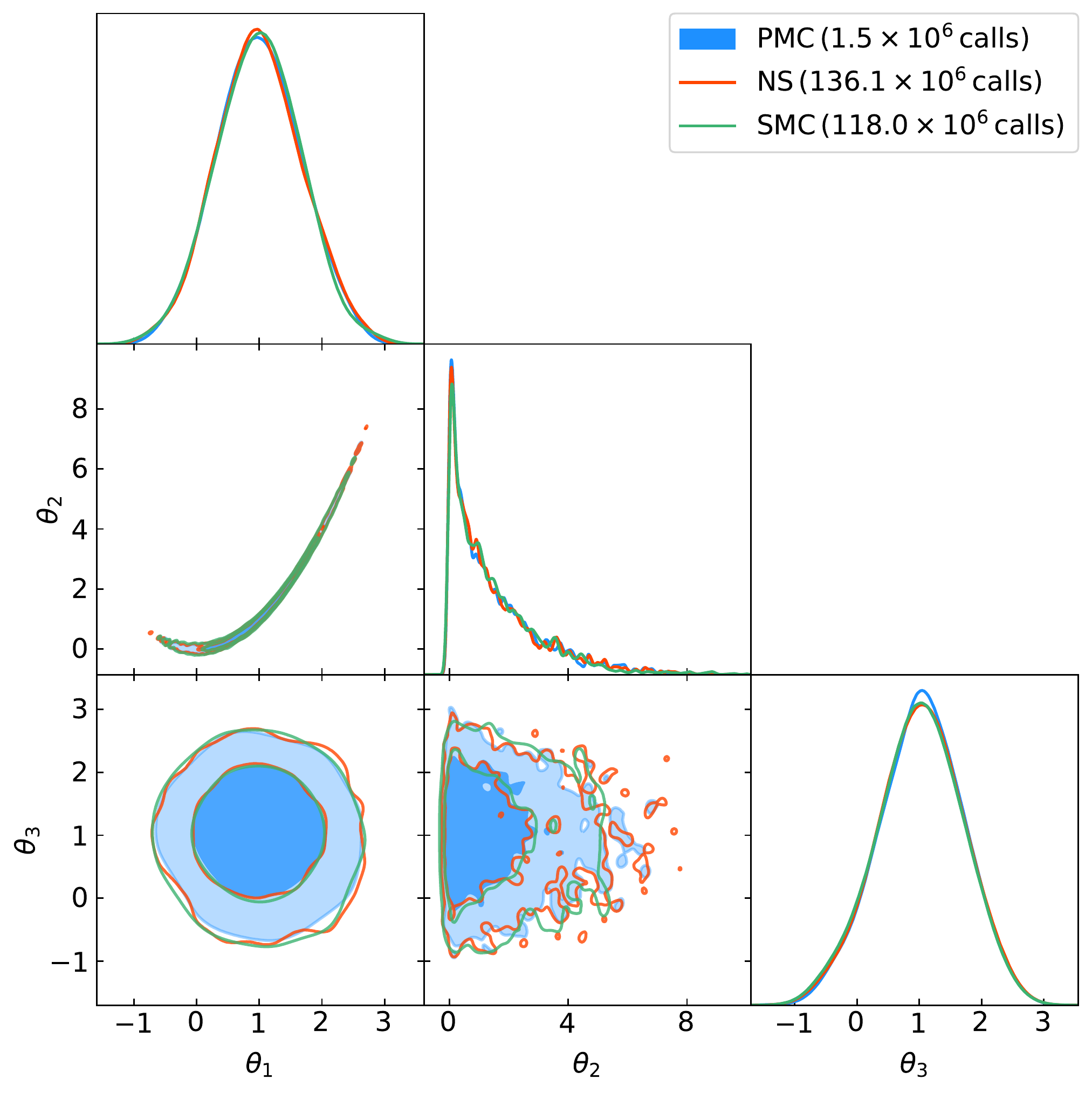}}
    \caption{Illustration of the 1--dimensional and 2--dimensional marginal posteriors for the first three out of $20$ parameters of the \emph{Rosenbrock} distribution. The figure shows the 1--$\sigma$ and 2--$\sigma$ contours generated by \emph{Preconditioned Monte Carlo} (PMC) in \emph{blue}, \emph{Nested Sampling} (NS) in \emph{orange}, and \emph{Sequential Monte Carlo} (SMC) in \emph{green}. The legend also shows the computational cost of each method in terms of the total number of required model evaluations until convergence is reached.}
    \label{fig:rosenbrock}
\end{figure}

The first toy example that we used is the \emph{Rosenbrock} distribution, which exhibits strong non--linear correlation between its parameters. For this reason, the \emph{Rosenbrock} distribution has often been used as a benchmark target for optimization and sampling tasks. Here we use a 20--dimensional generalisation defined through the probability density function given by:
\begin{equation}
    \label{eq:rosenbrock}
    \log P(\theta) = -\sum_{i=1}^{N/2}\left[10\left(\theta_{2i-1}^{2}-\theta_{2i}\right)^{2}+\left(\theta_{2i-1}-1\right)^{2} \right]\,.
\end{equation}
We use flat priors $\mathcal{U}(-10,10)$ for all parameters. Figure \ref{fig:rosenbrock} shows the 2--dimensional marginal posterior for the first two parameters as generated by the three methods. The total computational cost of PMC, NS, and SMC is $1.5\times 10^{6}$, $136.1\times 10^{6}$, and $118.0\times 10^{6}$ model evaluations, respectively. PMC requires approximately $1/91$ of the number of model evaluations that NS does and approximately $1/79$ of those that SMC does. 

\begin{table}
    \centering
    \caption{The table shows a comparison of PMC, NS, and SMC in terms of their computational cost (i.e. total number of model evaluations until convergence).}
    \def\arraystretch{1.1}
    \begin{tabular}{lcccc}
        \toprule[0.75pt]
        & & \multicolumn{3}{c}{Model evaluations $(\times 10^{6})$} \\
        \cmidrule(lr){3-5}
        Distribution & & \textbf{PMC} & NS & SMC \\
        \midrule[0.5pt]
        Rosenbrock & & $\mathbf{1.5}$ & $136.1$ & $118.0$  \\
        Gaussian Mixture & & $\mathbf{1.6}$ & $222.1$ & $9.6$  \\
        Primordial Features & & $\mathbf{0.4}$ & $21.3$ & $19.5$  \\
        Gravitational Waves &  & $\mathbf{0.4}$ & $10.2$ & $4.6$  \\
        \bottomrule[0.75pt]
    \end{tabular}
    \label{tab:comparison}
\end{table}

\subsection{Gaussian Mixture}

\begin{figure}
    \centering
	\centerline{\includegraphics[scale=0.42]{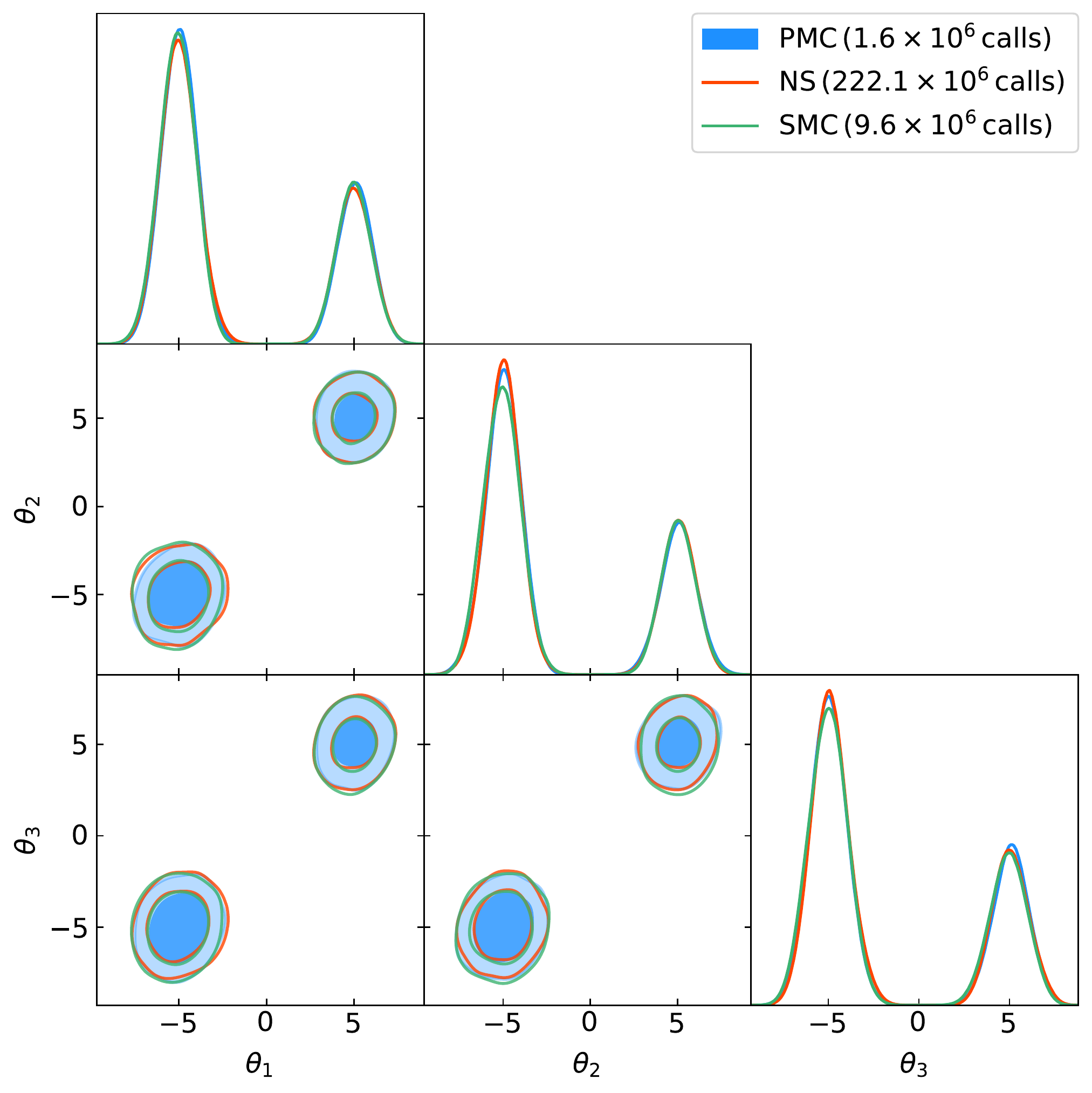}}
    \caption{Illustration of the 1--dimensional and 2--dimensional marginal posteriors for the first three out of $50$ parameters of the two--component Gaussian mixture distribution. The figure shows the 1--$\sigma$ and 2--$\sigma$ contours generated by \emph{Preconditioned Monte Carlo} (PMC) in \emph{blue}, \emph{Nested Sampling} (NS) in \emph{orange}, and \emph{Sequential Monte Carlo} (SMC) in \emph{green}. The legend also shows the computational cost of each method in terms of the total number of required model evaluations until convergence is reached.}
    \label{fig:mixture}
\end{figure}

The second toy example that we used is a 50--dimensional Gaussian Mixture with two components, one of them being twice as massive as the other. This is a highly multimodal problem as the target distribution exhibits two distinct modes that are well separated. Just as in the \emph{Rosenbrock} case, we use flat priors $\mathcal{U}(-10,10)$ for all parameters. Figure \ref{fig:rosenbrock} shows the 1--dimensional and 2--dimensional marginal posteriors for the first three parameters as generated by the three methods. The total computational cost of PMC, NS, and SMC is $1.6\times 10^{6}$, $222.1\times 10^{6}$, and $9.6\times 10^{6}$ model evaluations respectively. PMC requires approximately $1/139$ of the number of model evaluations that NS does and $1/6$ of those that SMC does.

\subsection{Primordial Features}

\begin{figure*}
    \centering
	\centerline{\includegraphics[scale=0.157]{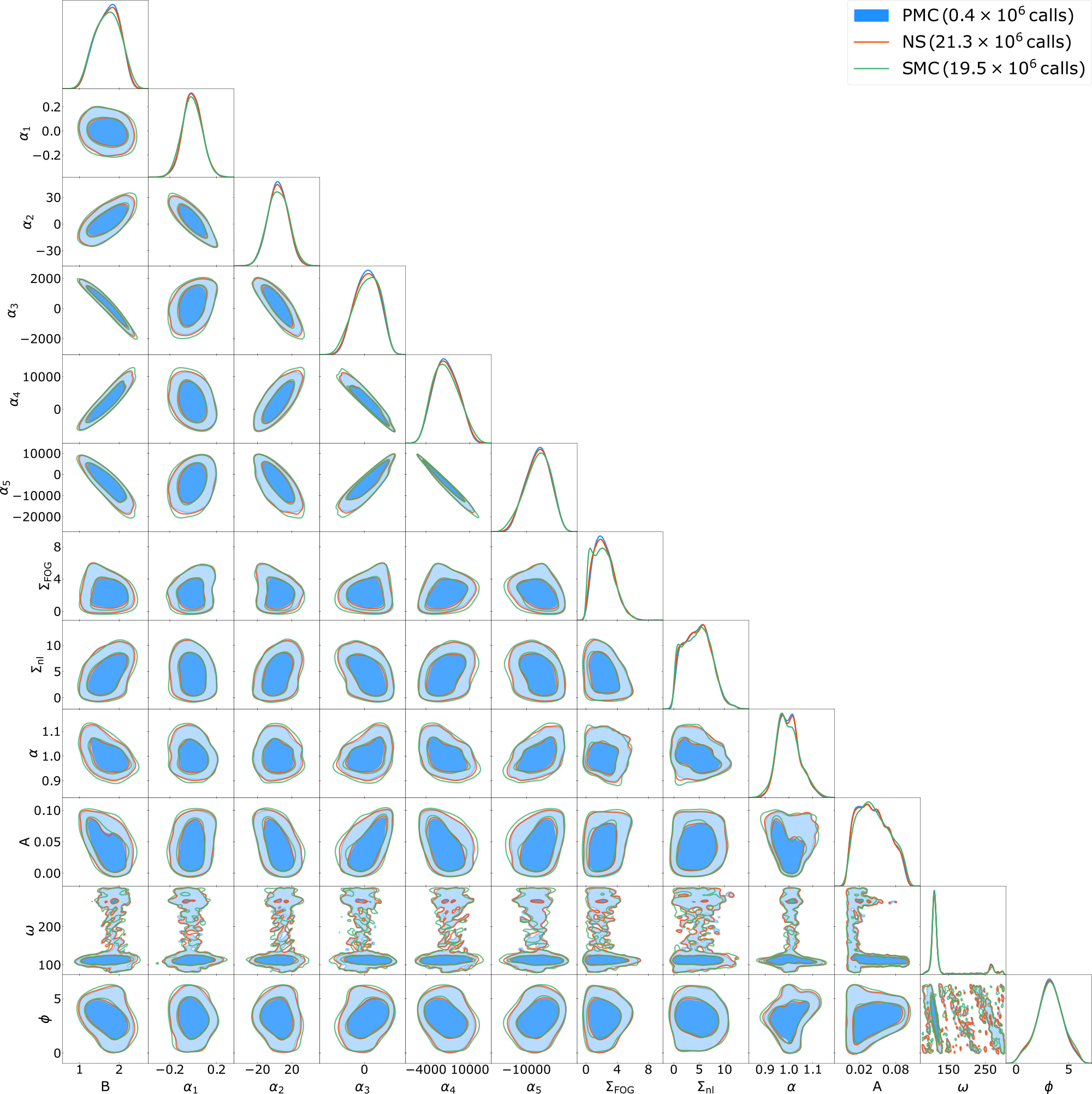}}
    \caption{Illustration of the 1--dimensional and 2--dimensional marginal posteriors for the $12$ parameters of the primordial features posterior. The figure shows the 1--$\sigma$ and 2--$\sigma$ contours generated by \emph{Preconditioned Monte Carlo} (PMC) in \emph{blue}, \emph{Nested Sampling} (NS) in \emph{orange}, and \emph{Sequential Monte Carlo} (SMC) in \emph{green}. The legend also shows the computational cost of each method in terms of the total number of required model evaluations until convergence is reached.}
    \label{fig:feature}
\end{figure*}

The first realistic application that we study is the search for primordial features along the Baryon Accoustic Oscillation (BAO) signature in the distribution of galaxies observed by the Sloan Digital Sky Survey (SDSS)~\citep{SDSSIII}. In particular, the data that we analysed come from the 12th data release (DR12) of the high--redshift North Galactic Cap (NGC) sample of the Baryon Oscillation Spectroscopic Survey (BOSS)~\citep{BOSS}. Our analysis follows closely that of ~\citet{beutler2019primordial} for the linear oscillation model. The inference problem includes $12$ free parameters with either flat/uniform or normal priors. Figure \ref{fig:feature} shows the 1--dimensional and 2--dimensional marginal posteriors of the aforementioned analysis. The posterior distribution exhibits a highly non--Gaussian geometry that can hinder the sampling performance of conventional methods. The total computational cost of PMC, NS, and SMC is $0.4\times 10^{6}$, $21.3\times 10^{6}$, and $19.5\times 10^{6}$ model evaluations respectively. PMC requires approximately $1/53$ of the number of model evaluations that NS does, and $1/49$ of those that SMC does.

\subsection{Gravitational Waves}

\begin{figure*}
    \centering
	\centerline{\includegraphics[scale=0.145]{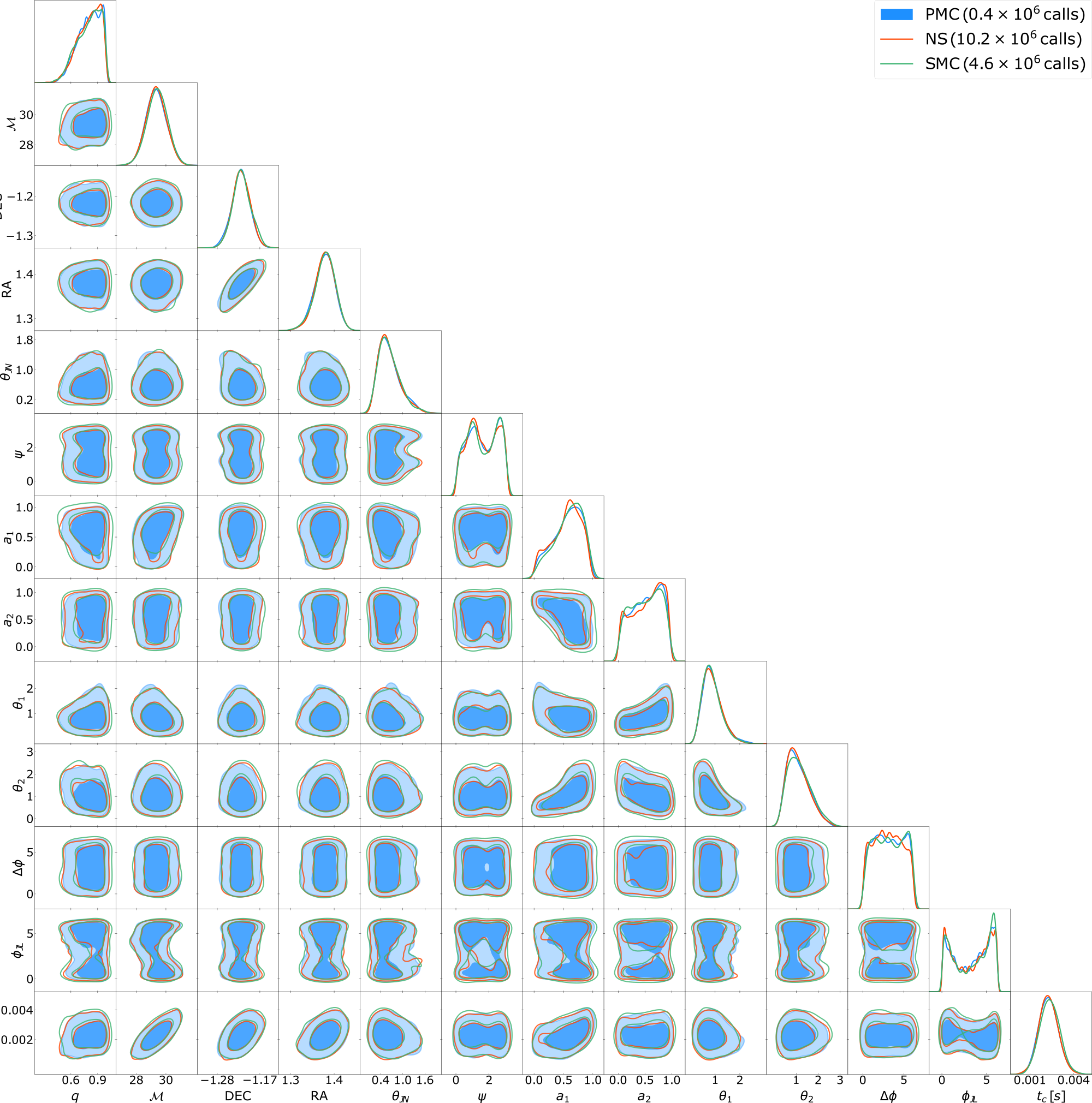}}
    \caption{Illustration of the 1--dimensional and 2--dimensional marginal posteriors for the $13$ parameters of the gravitational waves posterior. The figure shows the 1--$\sigma$ and 2--$\sigma$ contours generated by \emph{Preconditioned Monte Carlo} (PMC) in \emph{blue}, \emph{Nested Sampling} (NS) in \emph{orange}, and \emph{Sequential Monte Carlo} (SMC) in \emph{green}. The legend also shows the computational cost of each method in terms of the total number of required model evaluations until convergence is reached.}
    \label{fig:gw}
\end{figure*}

The second realistic application is the simulated gravitational wave analysis of an injected signal. For this, we used the standard CBC (i.e. compact binary coalescence) injected signal configuration provided by \texttt{BILBY}~\citep{ashton2019bilby}. The inference problem includes $13$ free parameters with a variety of common priors. Figure \ref{fig:gw} shows the 1--dimensional and 2--dimensional marginal posteriors of the aforementioned analysis. The posterior distribution exhibits a highly non--Gaussian geometry that can hinder the sampling performance of conventional methods. The total computational cost of PMC, NS, and SMC is $0.4\times 10^{6}$, $10.2\times 10^{6}$, and $4.6\times 10^{6}$ model evaluations respectively. PMC requires approximately $1/25$ of the number of model evaluations that NS does and $1/11$ of those that SMC does.

\section{Discussion}
\label{sec:discussion}

While we have demonstrated PMC's superior sampling performance for a number of target distributions, including two real--world applications, the real test is based on researchers applying the method to their analyses. Different applications pose different computational challenges and there is no one single sampler to rule them all. Sometimes, certain kinds of distributions will be better handled by other, perhaps simpler, approaches.

In general, we expect PMC to be a useful tool when dealing with computationally expensive likelihood functions and highly correlated or multimodal posteriors. There are two main reasons for this. First, training of the normalising flow takes about $\mathcal{O}(1\, {\rm s})$ per iteration on a laptop computer, whereas the actual vectorised evaluation of the bijective mapping takes almost $\mathcal{O}(10\, {\rm ms})$ per MCMC step for the whole population of particles. This means that if the cost of evaluating the likelihood is low enough to be comparable to that of the normalising flow, as discussed above, the chances are that there are simpler methods (e.g. MCMC) that can obtain the results more quickly. The second reason has to do with the geometry of the posterior distribution. If the latter is trivial enough (e.g. approximately Gaussian with no non--linear correlation or multiple modes), then the use of the normalising flow as a preconditioner would offer no benefit and instead only help delay the run.

On the other hand, if both of these conditions are met, that is, the likelihood function is computationally expensive, as is often the case in cosmology, and the posterior is non--Gaussian, then PMC can be a valuable asset in the astronomer's toolkit. Furthermore, when the cost of evaluating the likelihood function is large enough to dominate both the normalising flow evaluation and any potential \textit{MPI} communication overhead, one can capitalise on the availability of multiple CPUs in order to accelerate PMC. In particular, if the evaluation of the likelihood function takes $\mathcal{O}(1\, {\rm s})$, one should be able to use up to thousands of CPUs, potentially parallelising all or a substantial fraction of the particles.

\section{Conclusions}
\label{sec:conclusions}

We introduced PMC, a preconditioned generalisation of the standard SMC algorithm. PMC is a novel sampling method that can accelerate Bayesian inference and model comparison in computationally challenging astronomical and cosmological analyses.

After introducing the method in Section \ref{sec:method}, we presented a thorough demonstration of \textit{Preconditioned Monte Carlo}'s sampling capabilities by comparing its sampling performance to that of \textit{Nested Sampling} and \textit{Sequential Monte Carlo} in a range of  target distributions characterised by non--trivial geometry. The results are presented in Table \ref{tab:comparison}. We found that \textit{Preconditioned Monte Carlo} is one to two orders of magnitude faster than either \textit{Nested Sampling} or \textit{Sequential Monte Carlo}, both of which performed similarly to each other. Furthermore, in the realistic analyses of primordial features and gravitational waves, \textit{Preconditioned Monte Carlo} required approximately $50$ and $25$ times fewer model evaluations compared to NS in order to converge. The reduced computational cost, combined with the superior parallelisation scaling, renders \textit{Preconditioned Monte Carlo} ideal for astronomical and cosmological Bayesian analyses with computationally expensive, strongly correlated, multimodal and high--dimensional posteriors. 

We hope that \textit{Preconditioned Monte Carlo} will prove useful to the astronomical community by facilitating challenging Bayesian data analyses and enabling the investigation of complex models and sparse datasets. We also released a \texttt{Python} implementation of \textit{Preconditioned Monte Carlo}, called \texttt{pocoMC}, which is publicly available at \url{https://github.com/minaskar/pocomc} and detailed documentation with installation instructions and examples at \url{https://pocomc.readthedocs.io}.

\section*{Acknowledgements}

The authors extend their gratitude to Jamie Donald--McCann, Richard Grumitt, Biwei Dai, and James Sullivan for providing useful comments. MK would also like to thank George Vretinaris for providing valuable feedback on an early version of the code. This work has benefited from a variety of \texttt{Python} packages including \texttt{numpy}~\citep{van2011numpy}, \texttt{scipy}~\citep{virtanen2020scipy}, \texttt{torch}~\citep{paszke2019pytorch}, \texttt{matplotlib}~\citep{hunter2007matplotlib}, \texttt{seaborn}~\citep{Waskom2021seaborn}, \texttt{getdist}~\citep{lewis2019getdist}, \texttt{sklearn}~\citep{pedregosa2011scikit}, \texttt{tqdm}~\citep{da2019tqdm}, \texttt{dynesty}~\citep{speagle2020dynesty}, and \texttt{mpi4py}~\citep{dalcin2011parallel}. This project has received funding from the European Research Council (ERC) under the European Union's Horizon 2020 research and innovation program (grant agreement 853291), and  by the U.S. Department of Energy, Office of Science, Office of Advanced Scientific Computing Research under Contract No. DE-AC02-05CH11231 at Lawrence Berkeley National Laboratory to enable research for Data-intensive Machine Learning and Analysis. FB is a University Research Fellow.

\section*{Data Availability}

All data used in this work are publicly available. Power spectrum estimates, covariance matrices and window functions used in the cosmological inference example are available at \url{http:// www.sdss3.org/science/boss_publications.php}. 




\bibliographystyle{mnras}
\bibliography{references} 




\appendix
\section{Comparison to Independent Metropolis--Hastings Sequential Monte Carlo}

Recent practice in the literature \citep{albergo2019flow,williams2021nested, arbel2021annealed} is to use normalising flows as auxiliary densities for \textit{Importance Sampling (IS)} and \textit{Independent Metropolis--Hastings (IMH)} estimators. The latter approach can also be accommodated in the context of \textit{Sequential Monte Carlo (SMC)} as an alternative to PMC. For this reason we will offer an experimental comparison of PMC to IMH--SMC.

The IMH--SMC allgorithm is identical to \Algo{pmc} with the exception that the mutation step of line $12$ takes place using the modified \textit{Metropolis acceptance criterion}
\begin{equation}
    \label{eq:Metropolis_criterion_transform_independent}
    \alpha = \min \left( 1, \frac{p_{\theta}(f^{-1}(u'))q(u)\Big|\det \frac{\partial f^{-1}(u')}{\partial u'}\Big|}{p_{\theta }(f^{-1}(u))q(u ')\Big|\det \frac{\partial f^{-1}(u)}{\partial u}\Big|}\right)\,,
\end{equation}
instead of equation \ref{eq:Metropolis_criterion_transform}. The difference between the two criteria is that the proposal distribution $q(u)=\mathcal{N}(u\vert 0,1)$ is no longer conditional on the previous state of the Markov chain.

The number $M$ of IMH steps performed in each iteration of IMH--SMC is determined adaptively during the run, based on the observed acceptance rate $\alpha$, as
\begin{equation}
    \label{eq:number_of_imh_steps}
    M = \frac{\log (1 - p)}{\log (1 - \alpha)}\,,
\end{equation}
where $p$ is the target probability of generating a new independent sample. In our examples below, the value of $p$ is chosen such that the computational cost of IMH--SMC is similar to that of PMC for the same example. This results in $p>0.99$ which corresponds to very conservative sampling. 

\begin{figure}
    \centering
	\centerline{\includegraphics[scale=0.45]{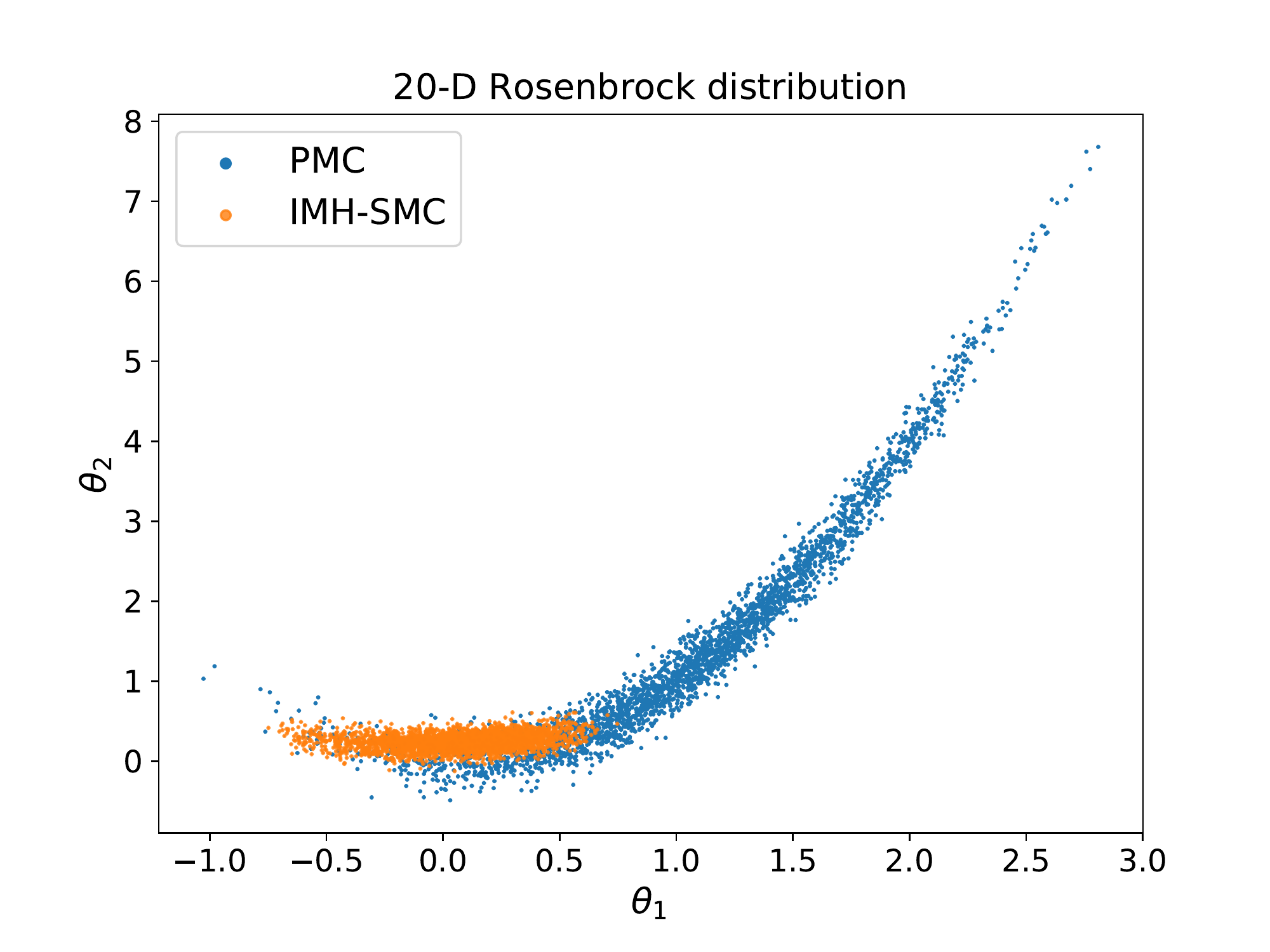}}
    \caption{Comparison of the first two parameters of samples generated using PMC (\textit{blue}) and IMH--SMC (\textit{orange}) for the $20$--D Rosenbrock target distribution. PMC produces representative samples, whereas IMH--SMC does not.}
    \label{fig:rosenbrock_comparison}
\end{figure}
\begin{figure}
    \centering
	\centerline{\includegraphics[scale=0.45]{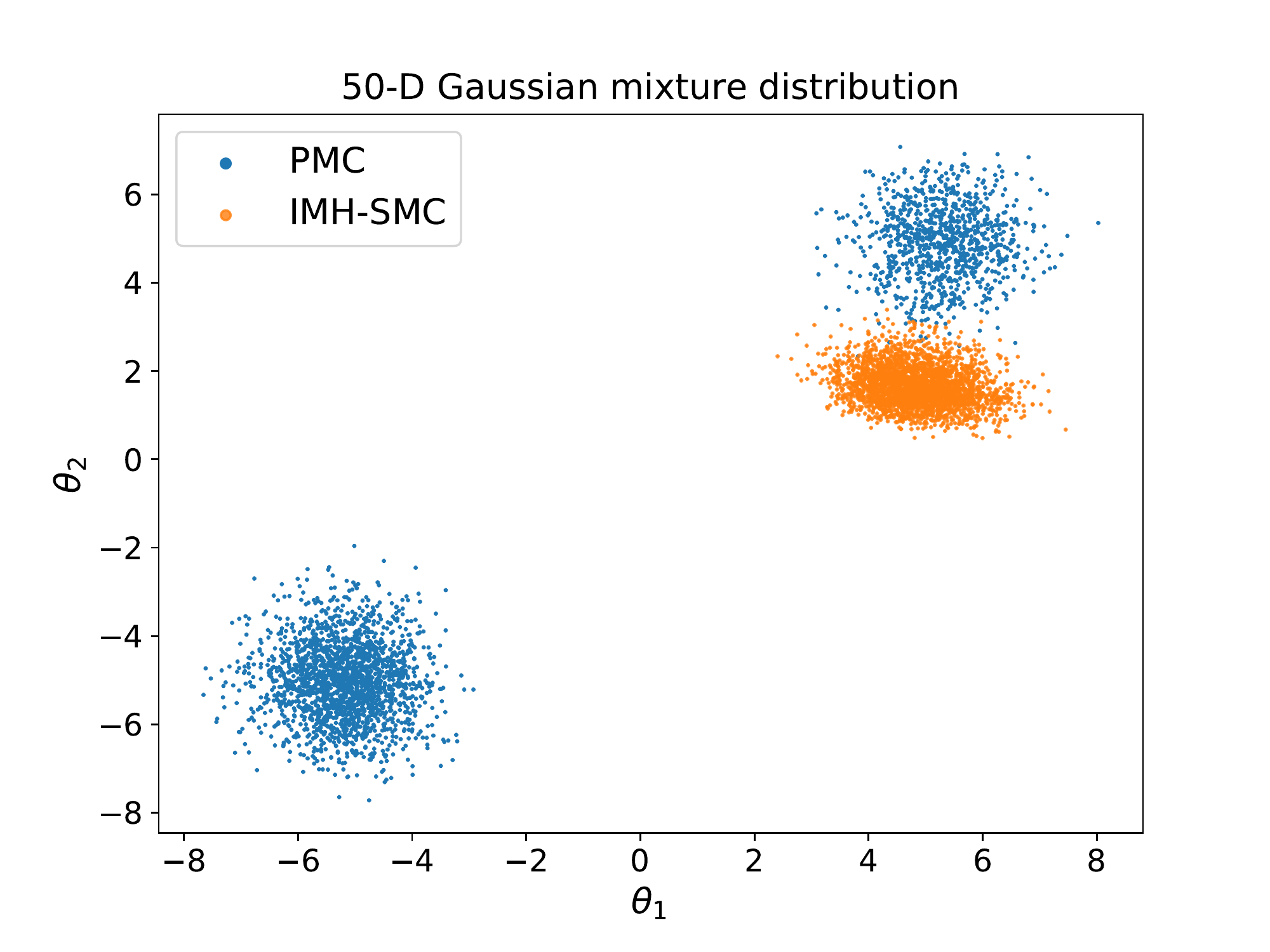}}
    \caption{Comparison of the first two parameters of samples generated using PMC (\textit{blue}) and IMH--SMC (\textit{orange}) for the $50$--D two--component Gaussian mixture target distribution. PMC produces representative samples, whereas IMH--SMC does not.}
    \label{fig:gaussian_mixture_comparison}
\end{figure}

Despite this, as shown in Figures \ref{fig:rosenbrock_comparison} and \ref{fig:gaussian_mixture_comparison}, for the $20$--dimensional Rosenbrock and the $50$--dimensional two--component Gaussian mixture studied in the main text respectively, IMH--SMC does not manage to produce typical samples from the posterior distribution. It is important to note here that the acceptance rate of IMH--SMC was high throughout both runs, and as such offered no indication on its own that NF is not correct.

The origin of this discrepancy between IMH--SMC and PMC in both cases, and the ultimate inability of IMH--SMC to compete with PMC, originates in the substantial mismatch between the NF distribution and target distribution in high dimensions and the subsequent over--fitting of the NF to the particle distribution leading to a narrower distribution. The high acceptance rate does not imply high quality of NF solution, and other tests of the quality of solution are needed, such as comparing expectation of $\log p$ between samples from NF and true MCMC samples. On the other hand, PMC does not suffer from this pathology as the local exploration offered by MCMC helps diversify the particles in order to avoid over--fitting. Furthermore, local MCMC methods generally scale better with the number of dimensions compared to IMH and IS.


\bsp	
\label{lastpage}
\end{document}